\documentclass[prl,twocolumn,superscriptaddress,floatfix,amsmath,nofootinbib,amssymb]{revtex4-1}
\usepackage{amssymb}
\usepackage{graphicx}
\usepackage{amsfonts}
\usepackage{amssymb}
\usepackage{natbib}

\begin{document}
\newcommand{\be}{\begin{eqnarray}}
\newcommand{\ee}{\end{eqnarray}}
\newcommand{\bea}{\begin{eqnarray}}
\newcommand{\eea}{\end{eqnarray}}
\newcommand{\bma}{\begin{subequations}}
\newcommand{\ema}{\end{subequations}}
\newcommand{\ket}[1]{\ensuremath{\left|{#1}\right\rangle}}
\newcommand{\bra}[1]{\ensuremath{\left\langle{#1}\right|}}
\def\RR{\mathbb{R}}
\def\E{\mathbf e}
\def\D{\boldsymbol \delta}
\def\dd{\delta}
\def\one{{\bf 1}}
\def\qq{{\rm q}}
\def\LL{{\rm L}}
\def\cc{{\rm c}}
\def\hh{{\rm h}}
\def\TTr{{\rm Tr}}

\title{Simulating accelerated atoms coupled to a quantum field}

\author{Marco del Rey}
\affiliation{Instituto de F\'{i}sica Fundamental, CSIC, Serrano 113-B, 28006 Madrid, Spain}

\author{Diego Porras}
\affiliation{Departamento de F\'isica Te\'orica I, Universidad Complutense, 28040 madrid, Spain}

\author{Eduardo Mart\'{i}n-Mart\'{i}nez}
\affiliation{Instituto de F\'{i}sica Fundamental, CSIC, Serrano 113-B, 28006 Madrid, Spain}

\date{\today}

\begin{abstract}
We show an analogy between static quantum emitters coupled to a single mode of a quantum field and accelerated Unruh-DeWitt detectors. We envision a way to simulate a variety of relativistic quantum field settings beyond the reach of current computational power, such as high number of qubits coupled to a quantum field following arbitrary non-inertial trajectories. Our scheme may be implemented with trapped ions and circuit QED set-ups.
\end{abstract}

\maketitle

{\it Introduction.--}
%
The study of accelerated atoms interacting with a quantum field is a fundamental problem, which has attracted a great deal of attention in General Relativity. Even though it has been treated extensively in the literature \cite{Crispino,Higuchi}, it still poses intriguing questions, as well as experimental challenges for the detection of quantum effects induced by acceleration. 
Moreover, the emerging field of relativistic quantum information has recently increased the attention drawn to the topic, and in particular the study of correlations and entanglement in non-inertial scenarios 
\cite{Vacbell}. 

Acceleration effects in quantum field theory are well understood in the perturbative regime, where, for example, transition rates are obtained for the excitation probability of atoms due to general relativity effects. 
A physical paradigm in this regime is the detection of the Unruh effect \cite{Unruh0}, which, roughly, implies that an atom accelerated in the vacuum of the field is excited in the same way as an inertial atom in an effective thermal field state. 
Going beyond the perturbative regime poses a tough theoretical problem, which gets even harder if one considers a set of emitters, in which case we face a many-body situation, situation particularly relevant from the point of view of relativistic quantum information.

In this letter we show that a system of accelerated atoms \cite{DeWitt} 
coupled to a bosonic field in the discrete mode approximation \cite{Berry,IvDragan}, shares interesting analogies to time-dependent problems in quantum optics.
Our work is motivated by the recent experimental progress that has allowed physicists to develop tools to control the dynamics of single emitters coupled to fields in set-ups such as trapped ions \cite{Leibfried} or circuit QED \cite{Wallraff04} . The application of those systems as analog-simulators of accelerated atoms is particularly relevant to many-body non-perturbative regimes, where numerical calculations are difficult. 
Furthermore, the insight gained by such analogies motivates the study of physical effects that may be relevant to the experimental detection of the Unruh effect. 
For example, collective phenomena such as superradiance \cite{Haroche} (which are known to amplify the effective coupling of a set of emitters to the field) could be used to increase detection sensitivity of quantum effects in non-inertial scenarios.

The letter is structured as follows: We start by presenting the Unruh-DeWitt detector model to characterize an atom coupled to a quantum field. Working in the interaction picture from the comoving atom reference frame, we show how to account for a uniform acceleration in the Hamiltonian. We then proceed to discuss the possible physical implementations in trapped ions and circuit QED. For both cases a model of emitters with controlled time-dependent atom-field couplings is presented, which, after some approximations, yields an identical Hamiltonian. For the sake of simplicity we will refer generically to single quantum emitters as ``atoms'', and the bosonic mode as ``field'', being the latter a phonon mode in a Coulomb crystal of trapped ions, or a photon mode confined in cavity in circuit QED.
Then, as an illustrative example,
we analyze the physics in the case of a single atom to predict the outcome of simple experimental realizations of our ideas.
An analogy to a decoupling process is explained in terms of the well-known Landau-Zener formula. Some future research lines on the topic will naturally emerge from our discussion.

{\it Accelerated Unruh DeWitt detectors.--} The Unruh-DeWitt detector \cite{DeWitt,LoSch,Crispino} is a standard model for a two-level atom coupled to a scalar field. This kind of detectors has been extensively used for multiple purposes such as acknowledging acceleration effects in cavities, measuring quantum correlations between spatially separated regions of spacetime, detecting entanglement degradation due to the Unruh and Hawking effects and proposing set-ups to directly detect such effects.
In general, computing time evolution under such a Hamiltonian is a complex problem, and thus, the affordable calculations reduce to very simple scenarios where perturbation theory to the first or second order is adequate.

On the other hand, special interest have the cases where the detectors couple only to a discrete number of modes. Presented in \cite{Berry} and extended in \cite{IvDragan} such settings directly model the interesting case of accelerated atoms going through a cavity. Moreover, they also feature a way to directly measure the Unruh effect considering cavities which are leaky to a finite number of modes \cite{Berry}. 

Let us start by presenting the discretized Unruh-DeWitt interaction Hamiltonian consisting in a set of two-level atoms coupled to a scalar field. We assume a set of atoms that are accelerating with proper acceleration $a$. From the atoms perspective the Hamiltonian can be rewritten as
\begin{align}\label{hamilto}
H_\text{I}&=\sum_{jm} g_{jm} (\sigma_j^+ e^{i\Omega_j \tau}+\sigma_j^-e^{-i\Omega_j \tau})\\*
&(a_m^\dagger e^{i\left[\omega_m t(\tau,\xi) -k_mx(\tau,\xi)\right]}+a_m e^{-i\left[\omega_m t(\tau,\xi) -k_mx(\tau,\xi)\right]})\nonumber 
\end{align}
where $(\tau,\xi)$ are the proper space-time coordinates of the accelerated detectors, and $(t,x)$ are Minkowskian coordinates. The following relation holds,
\begin{equation}\label{change}
ct=\xi \sinh\left(a\tau/c\right),\qquad x=\xi\cosh\left(a\tau/c\right).
\end{equation}
Directly from \eqref{change} we see that for constant $\xi$ these coordinates describe hyperbolic trajectories in  space-time whose asymptote is the light cone. A uniformly accelerated observer whose proper coordinates are \eqref{change}, follows the trajectory of constant Rindler position $\xi= c^2/a$ \cite{gravitation}.


If all the detectors follow such a trajectory we can rewrite the Hamiltonian in a form which is suitable to find quantum optical analogs,
\begin{equation}\label{hamil3}
H_{\rm I} =\sum_{jm} g_{jm}
\big( \sigma^+_j e^{i \Omega_j \tau} \! \!+ {\rm H.c.} \big)
\big(a_m^\dagger e^{i \Phi_m(\tau)}  \!+ {\rm H.c.} \big),
\end{equation}
\quad\\[-4mm]
where $\tau$ is the atoms proper time, and
\begin{align}
\Phi_m \!(\tau) &= \! - k_m \xi e^{- a \tau/c} = \! -\frac{k_m c^2}{a} e^{- a \tau/c} = \! -\frac{\omega_m}{\alpha} e^{- \alpha \tau}, \label{field.phase}
\end{align}
where we have used the  relation $\omega_m = c k_m$, and defined the parameter $\alpha = a/c$.
We can readily see how we recover the non-relativistic limit by taking the limit of small acceleration, 
$a^\dagger_m e^{i \Phi_m(\tau)} \to 
a^\dagger_m e^{- i k_m c^2/a + i \omega_m \tau -\frac{1}{2}k_m a \tau^2}$.
Note that the time-independent phase factor, $- k_m c^2/a$, may be simply absorbed in the definition of the particle operators. 
Eq. (\ref{hamil3}) defines the Hamiltonian of interest that we aim to simulate. It also reproduces a scenario where an array of detectors are resisting near the event horizon of a Schwarzschild black hole, if we consider in \eqref{hamil3} that $a \approx {\kappa}/{\sqrt{f_0}}$, 
where $\kappa$ is the surface gravity of the black hole and $f_0$ is the gravitational redshift factor, following the arguments in \cite{Edu6}. 

{\it Physical implementations. Trapped ions.--}
This setup is ideally suited to prepare and measure quantum states by well established experimental techniques \cite{Leibfried}, which have found an important application in quantum simulation of many-body physics \cite{Schneider11}, and single particle dynamics in special relativity \cite{naturekike}.  Previous proposals  also allowed the simulation of quantum fields using trapped ions in general relativistic settings but in different contexts \cite{Milburn2,Milburn}, however, paying the price of requiring control on the frequency of the qubit and the coupling strength. In our proposal,  only phase control is needed. Our setting is therefore readily exportable to other experimental set-ups, as shown below.

In our scheme we consider a chain of $N$ ions of mass $M$. For simplicity, we focus on the single-mode version of (\ref{hamil3}), and consider that the bosonic mode is a phonon mode of the chain, namely, the center-of-mass mode which accounts to a homogeneous displacement of all the ions. State-of-the-art techniques may be used to implement phonon sidebands, and to control the atom-field coupling.

The vibrations of a Coulomb chain consists of a set of normal modes
described by $H_0 = \sum_n \omega_n a^\dagger_n a_n$, with $\omega_n$ the normal mode frequencies, and $a_n$, $a_n^\dagger$, phonon operators. 
Levels $| 0 \rangle$ and $| 1 \rangle$ are two electronic states of the ions.
Two sets of lasers couple those levels by means of Raman transitions, with amplitudes $\Omega_{\LL, j}$, and two frequencies $\omega_{\LL_{1,2}}$, (see \cite{Leibfried} for details),
\begin{equation}
H_{\rm L}(\tau) =  \hspace{-0.2cm}
\sum_{j, \nu = 1,2} \hspace{-0.2cm}
\frac{\Omega_{\LL,j}}{2} \sigma^+_j 
e^{i k_\LL x_j} e^{- i \omega_{{\rm L}_\nu} \tau - i \phi_\nu(\tau)}
+ {\rm H.c.} .
\label{ion.laser}
\end{equation}
This equation  represents a standard atom-light interaction term, with the only peculiarity that we consider time dependent phases $\phi_1(\tau)$, $\phi_2(\tau)$.
$x_j$ are operators corresponding to the ion displacements, relative to the equilibrium positions along the chain.
We express those displacements in terms of phonon operators, 
$x_j = \sum_n {\cal M}_{j,n} \bar{x}_n (a_n + a_n^\dagger)$, where ${\cal M}_{j,n}$ are phonon wavefunctions, and ${\bar{x}}_n = 1/\sqrt{2 M \omega_n}$. 

We choose laser frequencies close to resonance with the center-of-mass mode, $n = 0$, such that 
$\omega_{{\rm L}_1} = - \omega_0 - \Omega$, 
$\omega_{{\rm L}_2} =  \omega_0 - \Omega$, with $\Omega \ll \omega_0$. The coupling (\ref{ion.laser}) simulates the quantum dynamics of an accelerated Unruh-DeWitt detector if: 
(i) $k \bar{x}_n \ll 1$ (Lamb-Dicke regime), such that 
we can expand the exponential in powers of $\delta x_j$, and restrict only to  linear atom-field couplings.
(ii) 
$\Omega_{\LL,j}/2 \ll \omega_0$, and $(\Omega_{\LL,j} / 2) k \bar{x}_n   \ll \omega_0, |\omega_0 - \omega_{n \neq 0}|$, so that we can neglect, in a rotating wave approximation (r.w.a.), all couplings to vibrational modes $n \neq 0$. This yields
\begin{align*}
H_{\rm L}(\tau) &=\sum_j g_j  \sigma^+_j e^{i \Omega \tau}  \big(a_0 e^{- i \phi_1 (\tau)}+a_0^\dagger e^{- i \phi_2 (\tau)}\big)  + {\rm H.c.},
\end{align*}
where we have used that the center-of-mass vibrational modes fulfills ${\cal M}_{j,0} = 1/ \sqrt{N}$, such  that $g_j = i\Omega_{\LL_j} k \bar{x}_0 / (2 \sqrt{N})$.  By choosing phases such that $\phi_1(\tau) = - \phi_2(\tau) = \Phi(\tau)$, we arrive at the single-mode version of Eq. (\ref{hamil3}). Including more modes would require additional lasers on resonance with other vibrational modes.
Position depending couplings, $g_j$, and frequencies, $\Omega_j$, may be achieved given a certain intensity profile of the lasers, and by focusing multiple lasers on each ion, respectively.

Typical experimental values are $\omega_0, |\omega_n - \omega_0| \approx$ 1 MHz, and $k \bar{x}_0 \approx$ $0.2$, such that our requirements are fulfilled with values $\Omega_\LL/2 =$ $100$  kHz, and $\Omega =$ $100$ kHz, yielding $g = 20/\sqrt{N}$ kHz.
Those values are well above typical decoherence rates in trapped ions. Also, in order to observe analogs to acceleration, $\Phi(\tau)$ has to increase exponentially over a time-scale comparable to the inverse energies involved in the set-up. For example, values $\alpha = 10^{-3} \Omega =$ $0.1$ kHz, would require to vary the optical phase on times scales of $10$ ms. This is technically feasible by using, for example, acousto-optical modulators, and standard experimental techniques from trapped ion quantum computation, which indeed require manipulation on a much shorter time-scale \cite{Leibfried03}. Our ideas may also be used with spin many-boson models as in \cite{Porras08}.

{\it Physical implementations. Circuit QED.--}
Our second proposed implementation consists of a superconducting qubit coupled to a microwave cavity in the strong-coupling regime \cite{Wallraff04}. 
We present a scheme that relies on the driving of the qubit frequency only, and is particularly well suited for our work \cite{Porras11}. 
The qubits and field noninteracting Hamiltonian is
\begin{equation}
H_{0}(\tau) = \omega_0 a^\dagger a + \frac{\epsilon}{2} \sum_j \sigma_{z,j} + H_{\rm d}(\tau) ,
\end{equation}
\quad\\[-4mm]
where $\omega$ is the resonant frequency of the mode, and $\epsilon$ is the qubit energy. $H_{\rm d}(\tau)$ describes a driving field, for which we assume the following form,
\begin{equation}
H_{\rm d}(\tau) 
= - \sum_j \sum_{\nu = 1,2} 
\eta_j \ \omega_{{\rm d}_{\nu,j}} 
\cos(\omega_{{\rm d}_{\nu,j}} t + \phi_\nu(\tau)) \sigma_{z,j}.
\label{driving.field}
\end{equation}
Note that Eq. \eqref{driving.field} is written as a periodic driving with a phase, $\phi_\nu (\tau)$, which can be considered to evolve slowly in time, in a sense to be quantified below.
The qubit-cavity coupling in the Schr\"odinger picture is given by
$H_{\rm I} = g_0 \left( \sigma^+ + \sigma^- \right) \left( a + a^\dagger \right)$.
We write the coupling in the interaction picture with respect to $H_0(\tau)$,
\begin{eqnarray}
H_{\rm I}(\tau) 
&=& g_0
\sum_j (\sigma_j^+ e^{i \epsilon \tau} {\cal G}_j(\tau) + {\rm H.c.})
(a e^{-i \omega_0 \tau} + {\rm H.c.}) ,
\nonumber \\
{\cal G}_j(\tau) &=& e^{- 2 i \sum_{\nu} \eta \sin(\omega_{d_{\nu,j}} \tau + \phi_\nu(\tau))} ,
\label{int.ham}
\end{eqnarray}
where we made the approximation that 
$\dot{\phi}(\tau)/\phi(\tau) \ll \omega_{{\rm d}_{\nu,j}}$.
Consider now that $\eta \ll 1$, and the choice of frequencies
$\omega_{{\rm d}_{1,j}} =$ $\epsilon - \Omega_j - \omega_0$, 
$\omega_{{\rm d}_{2,j}} =$ $\epsilon - \Omega_j + \omega_0$ ,
and conditions $\epsilon, \omega_0, |\epsilon - \omega_0| \gg g_0$. We expand 
${\cal G}(\tau)$ to first order in $\eta_j$, and keep only slow-rotating terms in a r.w.a:
\begin{equation*}
H_{\rm I}(\tau) \approx \eta g_0 
\sum_j \sigma^+_j (
e^{i \Omega_j \tau- i \phi_1(\tau)} a 
+ 
e^{i \Omega_j \tau - i \phi_2(\tau)} a^\dagger) + 
{\rm H.c.} .
\end{equation*}
\quad\\[-4mm]
This expression takes the form (\ref{hamil3}) by choosing $\phi_1(\tau) = \Phi(\tau)$, $\phi_2(\tau) = - \Phi(\tau)$, and $g_j = g_0 \eta_j$.

In circuit QED, the high energy scales $\epsilon$, $\omega_0$ are in the GHz regime. Low energy scales, such as $g_j$, $\Omega_j$, may be then in the MHz range. Finally, note that
$\dot{\phi}(\tau)/\phi(\tau) = \alpha$. Thus, in order to observe effects from effective acceleration, our scheme requires $\alpha$ on the frequency scale of $g_j$, $\Omega_j$, such that the driving fields in (\ref{driving.field}) have to be controlled with an inverse time in the MHz range. This seems technically feasible, since this rate is slower than typical evolution times in circuit-QED systems. By using different photonic modes in a cavity, as well as local control of qubit couplings, the full multi-mode Hamiltonian (\ref{hamil3}) may be implemented.

{\it Single detector case: non-adiabatic effects induced by acceleration.--} We will study a simple case with a two-folded purpose in mind: on the one hand it will serve to gain some insight on the parameters required for our experimental proposal while, on the other hand, it will unveil an analogy between non-equilibrium physics and quantum effects induced by acceleration. Let us consider a single atom $A$, with natural frequency $\Omega$ and uniform acceleration frequency $\alpha=a/c=f_a\Omega$, coupled to a single-mode field with proper frequency $\omega > \Omega$. 

Let us have $A$ excited at $\tau=0$ with no excitations in the field. Representing the  free eigenstates of atom ($A$) and field ($F$) with the notation $\ket{\psi}=\ket{A\,F}$, the initial state would then be $\ket{i}=\ket{e\,0}$. 
We let the system evolve naturally and concentrate on the probability for $A$ to decay to its ground state:
\begin{equation} \label{eq:final.probability}
\mathcal{P}_{A_g} (\tau)=\sum_F |\langle g\, F|U(\tau)|e\,0\rangle|^2.
\end{equation}

It is convenient to analyze the phase in the Hamiltonian (\ref{hamil3}). The latter is equivalent to a Hamiltonian in the interaction picture with respect to a time dependent effective frequency 
\begin{equation} \label{eq:effective.field}
\omega_{\text{eff}} (\tau)=\partial_\tau\Phi(\tau) = \omega_0 e^{- \alpha \tau},
 \end{equation}
which decreases asymptotically. 
We can therefore picture this situation as a system starting in a certain low energy state, in this case matching $|e\,0\rangle$, coupled to a high energy state which initially happens to be $|g\,1\rangle$. The difference between the two effective energies varies with time as
\begin{equation} \label{eq:effective.energy.difference}
\Delta E_{\text{eff}} (\tau)= \omega_0 e^{- \alpha \tau}- \Omega,
 \end{equation}
and there will be a level crossing, corresponding to a stationary $\Phi(\tau)$, taking place at $\tau_c= \ln (\omega_0/\Omega)/\alpha$
\begin{figure}[h]
\begin{center}
\includegraphics[width=0.50\textwidth]{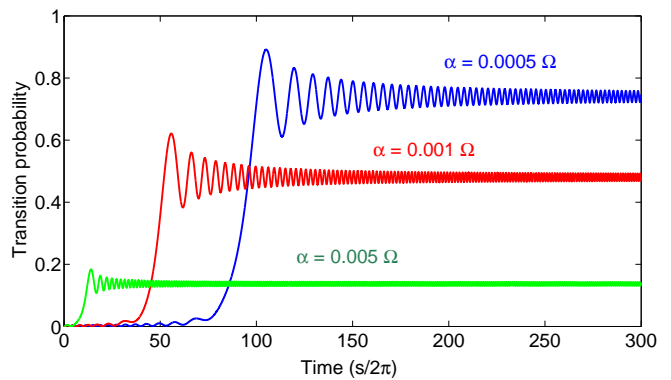}
\end{center}
\caption{(Color online) Time evolution of $\mathcal P_{A_g}$ for different accelerations, with the  parameter values $g=0.01\Omega$ and $\omega_0=1.33\Omega$. The crossing time and the transition probability fit the Landau-Zener prediction described in \eqref{eq:landau.zener}.} \label{fig:1}
\end{figure}
After a long time (as compared to $1/\alpha$) we can obviously approximate the low energy level of the system as $|g\,1\rangle$, and the high as $|e\,0\rangle$
We can therefore establish an analogy between this kind of phenomenon and a typical Landau-Zener transition, making therefore a link between non-equilibrium physics and quantum field theory in curved space-times.
If the energy difference happens to vary very slowly (which will be the case if $\alpha\ll g$), we would expect no transition between eigenstates to take place, so the system will stay in the ground state, which means actually decaying into  $|g\,1\rangle$.  If however, the evolution happens to be diabatic, the probability of transiting into a high excited state (so to say, staying in $|e\,0\rangle$) can be approximated by the Landau-Zener formula: 
\begin{equation} \label{eq:landau.zener}
\mathcal{P}_{{A_e}_{(+\infty)}} \simeq e^{-2\pi\Gamma}\Rightarrow \mathcal{P}_{{A_g}_{(+\infty)}} \simeq 1- e^{-2\pi\Gamma},
\end{equation}
where $\Gamma = g^2/(\partial_\tau \Delta E|_{\tau=\tau_c})$.
$\mathcal{P}_{{A_g}}$ is plotted in Fig. \ref{fig:1}.

\begin{figure}[h]
\begin{center}
\includegraphics[width=0.50\textwidth]{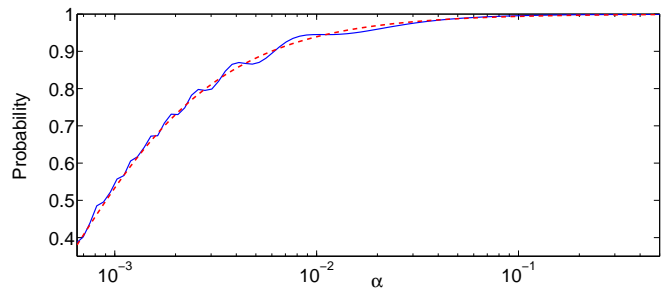}
\end{center}
\caption{(Color online) Comparison between Landau-Zener theoretical prediction (red dashed) and probability for an accelerated detector of remaining in the excited state as a function of the acceleration (blue solid) with parameter values $g = 0.01\Omega$ and $\omega_0=1.33 \Omega$.} \label{fig:2}
\end{figure}

We note several differences with  the original Landau-Zener model \cite{zener}. Namely, \eqref{eq:effective.energy.difference} is not linear but exponential in time; also, in our case, only for small couplings we have a two-level problem (a r.w.a. approximation cannot be performed). Nevertheless, in the regimes considered, L-Z is a very good approximation as shown in Fig. \ref{fig:2}
 The atom stays excited until the phase variation rate slows down at times close to $\tau_c$. Around that moment the coupling terms get resonant and the decay probability grows notably. From then on atom and field  get progressively decoupled again with the probability stabilizing itself.

{\it Conclusions.--}
We have presented a scheme for simulating a set of accelerated atoms coupled to a single-mode field. First, we have identified the Hamiltonian which yields the evolution as seen by a comoving observer with the atoms. A method has been presented to obtain the same Hamiltonian for circuit QED and trapped ions  by inducing time-dependent sidebands in atom-field couplings. 
Our idea may be extended to many-particle experiments which could simulate results that are not affordable for classical computers such as arbitrary non-inertial trajectories of detectors and many detectors coupled to quantum fields. 
Finally, by interpreting our results from a Landau-Zener perspective, we have made a connection between non-equilibrium Physics and quantum effects due to acceleration. This reveals an interesting point that will undoubtedly be exploited in the future: general relativistic quantum effects can be studied with tools coming from non-equilibrium Physics.
  
The authors thank M. Montero for his helpful comments,
and J.J. Garc\'{\i}a-Ripoll for discussions. 
E. M-M. and M. dR. were supported by a CSIC JAE-PREDOC grant. M.dR. was also supported by Residencia de Estudiantes. 
D. P. was supported by RyC Contract Y200200074 and EU (PICC), FIS2009-10061, CAM-UCM/910758. The authors were partially supported by QUITEMAD S2009-ESP-1594.

%


\begin{thebibliography}{20}%
\makeatletter
\providecommand \@ifxundefined [1]{%
 \@ifx{#1\undefined}
}%
\providecommand \@ifnum [1]{%
 \ifnum #1\expandafter \@firstoftwo
 \else \expandafter \@secondoftwo
 \fi
}%
\providecommand \@ifx [1]{%
 \ifx #1\expandafter \@firstoftwo
 \else \expandafter \@secondoftwo
 \fi
}%
\providecommand \natexlab [1]{#1}%
\providecommand \enquote  [1]{``#1''}%
\providecommand \bibnamefont  [1]{#1}%
\providecommand \bibfnamefont [1]{#1}%
\providecommand \citenamefont [1]{#1}%
\providecommand \href@noop [0]{\@secondoftwo}%
\providecommand \href [0]{\begingroup \@sanitize@url \@href}%
\providecommand \@href[1]{\@@startlink{#1}\@@href}%
\providecommand \@@href[1]{\endgroup#1\@@endlink}%
\providecommand \@sanitize@url [0]{\catcode `\\12\catcode `\$12\catcode
  `\&12\catcode `\#12\catcode `\^12\catcode `\_12\catcode `\%12\relax}%
\providecommand \@@startlink[1]{}%
\providecommand \@@endlink[0]{}%
\providecommand \url  [0]{\begingroup\@sanitize@url \@url }%
\providecommand \@url [1]{\endgroup\@href {#1}{\urlprefix }}%
\providecommand \urlprefix  [0]{URL }%
\providecommand \Eprint [0]{\href }%
\@ifxundefined \urlstyle {%
  \providecommand \doi  [0]{\begingroup \@sanitize@url \@doi}%
  \providecommand \@doi [1]{\endgroup \@@startlink {\doibase
  #1}doi:\discretionary {}{}{}#1\@@endlink }%
}{%
  \providecommand \doi  [0]{doi:\discretionary{}{}{}\begingroup
  \urlstyle{rm}\Url }%
}%
\providecommand \doibase [0]{http://dx.doi.org/}%
\providecommand \Doi [0]{\begingroup \@sanitize@url \@Doi }%
\providecommand \@Doi  [1]{\endgroup\@@startlink{\doibase#1}\@@Doi}%
\providecommand \@@Doi [1]{#1\@@endlink}%
\providecommand \selectlanguage [0]{\@gobble}%
\providecommand \bibinfo  [0]{\@secondoftwo}%
\providecommand \bibfield  [0]{\@secondoftwo}%
\providecommand \translation [1]{[#1]}%
\providecommand \BibitemOpen [0]{}%
\providecommand \bibitemStop [0]{}%
\providecommand \bibitemNoStop [0]{.\EOS\space}%
\providecommand \EOS [0]{\spacefactor3000\relax}%
\providecommand \BibitemShut  [1]{\csname bibitem#1\endcsname}%
\bibitem [{\citenamefont {Crispino}\ \emph {et~al.}(2008)\citenamefont
  {Crispino}, \citenamefont {Higuchi},\ and\ \citenamefont
  {Matsas}}]{Crispino}%
  \BibitemOpen
  \bibfield  {author} {\bibinfo {author} {\bibfnamefont {L.~C.~B.}\
  \bibnamefont {Crispino}}, \bibinfo {author} {\bibfnamefont {A.}~\bibnamefont
  {Higuchi}}, \ and\ \bibinfo {author} {\bibfnamefont {G.~E.~A.}\ \bibnamefont
  {Matsas}},\ }\href@noop {} {\bibfield  {journal} {\bibinfo  {journal} {Rev.
  Mod. Phys.},\ }\textbf {\bibinfo {volume} {80}},\ \bibinfo {pages} {787}
  (\bibinfo {year} {2008})}\BibitemShut {NoStop}%
\bibitem [{\citenamefont {Higuchi}\ \emph {et~al.}(1992)\citenamefont
  {Higuchi}, \citenamefont {Matsas},\ and\ \citenamefont {Sudarsky }}]{Higuchi}%
  \BibitemOpen
  \bibfield  {author} {\bibinfo {author} {\bibfnamefont {A.}~\bibnamefont
  {Higuchi}}, \bibinfo {author} {\bibfnamefont {G.~E.~A.}\ \bibnamefont
  {Matsas}}, \ and\ \bibinfo {author} {\bibfnamefont {D.}~\bibnamefont
  {Sudarsky}},\ }{\bibfield  {journal}
  {\bibinfo  {journal} {Phys. Rev. D},\ }\textbf {\bibinfo {volume} {46}},\
  \bibinfo {pages} {3450} (\bibinfo {year} {1992})}\BibitemShut {NoStop}%
\bibitem [{\citenamefont {Reznik}\ \emph {et~al.}(2005)\citenamefont {Reznik},
  \citenamefont {Retzker},\ and\ \citenamefont {Silman}}]{Vacbell}%
  \BibitemOpen
  \bibfield  {author} {\bibinfo {author} {\bibfnamefont {B.}~\bibnamefont
  {Reznik}}, \bibinfo {author} {\bibfnamefont {A.}~\bibnamefont {Retzker}}, \
  and\ \bibinfo {author} {\bibfnamefont {J.}~\bibnamefont {Silman}},\
  }\href@noop {} {\bibfield  {journal} {\bibinfo  {journal} {Phys. Rev. A}\
  }\textbf {\bibinfo {volume} {71}},\ \bibinfo {pages} {042104} (\bibinfo
  {year} {2005}),}\BibitemShut {NoStop}%
  \bibfield  {author} {\bibinfo {author} {\bibfnamefont {I.}~\bibnamefont
  {Fuentes-Schuller}}\ and\ \bibinfo {author} {\bibfnamefont {R.~B.}\
  \bibnamefont {Mann}},\ }\href@noop {} {\bibfield  {journal} {\bibinfo
  {journal} {Phys. Rev. Lett.}\ }\textbf {\bibinfo {volume} {95}},\ \bibinfo
  {pages} {120404} (\bibinfo {year} {2005}),}
  \bibfield  {author} {\bibinfo {author} {\bibfnamefont {S.-Y.}\ \bibnamefont
  {Lin}}\ and\ \bibinfo {author} {\bibfnamefont {B.}~\bibnamefont {Hu}},\
  }\href@noop {} {\bibfield  {journal} {\bibinfo  {journal} {Found. Phys.}\
  }\textbf {\bibinfo {volume} {37}},\ \bibinfo {pages} {480} (\bibinfo {year}
  {2007}),}
  \bibfield  {author} {\bibinfo {author} {\bibfnamefont {S.-Y.}\ \bibnamefont
  {Lin}}, \bibinfo {author} {\bibfnamefont {C.-H.}\ \bibnamefont {Chou}}, \
  and\ \bibinfo {author} {\bibfnamefont {B.~L.}\ \bibnamefont {Hu}},\
  }\href@noop {} {\bibfield  {journal} {\bibinfo  {journal} {Phys. Rev. D}\
  }\textbf {\bibinfo {volume} {78}},\ \bibinfo {pages} {125025} (\bibinfo
  {year} {2008}),}  \bibfield  {author} {\bibinfo {author} {\bibfnamefont {S.-Y.}\ \bibnamefont
  {Lin}}\ and\ \bibinfo {author} {\bibfnamefont {B.~L.}\ \bibnamefont {Hu}},\
  }\href@noop {} {\bibfield  {journal} {\bibinfo  {journal} {Classical and
  Quantum Gravity}\ }\textbf {\bibinfo {volume} {25}},\ \bibinfo {pages}
  {154004} (\bibinfo {year} {2008}),}
  \bibfield  {author} {\bibinfo {author} {\bibfnamefont {A.~G.~S.}\
  \bibnamefont {Landulfo}}, \bibinfo {author} {\bibfnamefont {G.~E.~A.}\
  \bibnamefont {Matsas}}, \ and\ \bibinfo {author} {\bibfnamefont {A.~C.}\
  \bibnamefont {Torres}},\ }\href {\doibase 10.1103/PhysRevA.81.044103}
  {\bibfield  {journal} {\bibinfo  {journal} {Phys. Rev. A}\ }\textbf {\bibinfo
  {volume} {81}},\ \bibinfo {pages} {044103} (\bibinfo {year}
  {2010}),}  \bibfield  {author} {\bibinfo {author} {\bibfnamefont {I.}~\bibnamefont
  {Fuentes}}, \bibinfo {author} {\bibfnamefont {R.~B.}\ \bibnamefont {Mann}},
  \bibinfo {author} {\bibfnamefont {E.}~\bibnamefont
  {Mart\'\i{}n-Mart\'\i{}nez}}, \ and\ \bibinfo {author} {\bibfnamefont
  {S.}~\bibnamefont {Moradi}},\ }\href@noop {} {\bibfield  {journal} {\bibinfo
  {journal} {Phys. Rev. D}\ }\textbf {\bibinfo {volume} {82}},\ \bibinfo
  {pages} {045030} (\bibinfo {year} {2010})}\BibitemShut {NoStop}%
\bibitem [{\citenamefont {Unruh}(1976)}]{Unruh0}%
  \BibitemOpen
  \bibfield  {author} {\bibinfo {author} {\bibfnamefont {W.~G.}\ \bibnamefont
  {Unruh}},\ }\href@noop {} {\bibfield  {journal} {\bibinfo  {journal} {Phys.
  Rev. D},\ }\textbf {\bibinfo {volume} {14}},\ \bibinfo {pages} {870}
  (\bibinfo {year} {1976})}\BibitemShut {NoStop}%
\bibitem [{\citenamefont {DeWitt}(1980)}]{DeWitt}%
  \BibitemOpen
  \bibfield  {author} {\bibinfo {author} {\bibnamefont {DeWitt}},\ }\href@noop
  {} {\emph {\bibinfo {title} {General Relativity; an Einstein Centenary
  Survey}}}\ (\bibinfo  {publisher} {Cambridge University Press (Cambridge,
  UK)},\ \bibinfo {year} {1980})\ ISBN \bibinfo {isbn} {0521299284}\BibitemShut
  {NoStop}%
\bibitem [{\citenamefont {Mart\'in-Mart\'inez}\ \emph
  {et~al.}(2011)\citenamefont {Mart\'in-Mart\'inez}, \citenamefont {Mann},\
  and\ \citenamefont {Fuentes}}]{Berry}%
  \BibitemOpen
  \bibfield  {author} {\bibinfo {author} {\bibfnamefont {E.}~\bibnamefont
  {Mart\'in-Mart\'inez}}, \bibinfo {author} {\bibfnamefont {R.~B.}\
  \bibnamefont {Mann}}, \ and\ \bibinfo {author} {\bibfnamefont
  {I.}~\bibnamefont {Fuentes}},\ }\href@noop {} {\bibfield  {journal} {\bibinfo
   {journal} {\emph{To appear in Phys. Rev. Lett.}}} (\bibinfo {year}
  {2011})},\ \Eprint {http://arxiv.org/abs/arXiv.org:1012.2208}
  {arXiv.org:1012.2208} \BibitemShut {NoStop}%
\bibitem [{\citenamefont {Dragan}\ and\ \citenamefont {Fuentes}()}]{IvDragan}%
  \BibitemOpen
  \bibfield  {author} {\bibinfo {author} {\bibfnamefont {A.}~\bibnamefont
  {Dragan}}\ and\ \bibinfo {author} {\bibfnamefont {I.}~\bibnamefont
  {Fuentes}},\ }\href@noop {} {\enquote {\bibinfo {title} {Probing the
  spacetime structure of vacuum entanglement},}\ }\Eprint
  {http://arxiv.org/abs/arXiv.org:1105.1192} {arXiv.org:1105.1192} \BibitemShut
  {NoStop}%
\bibitem [{\citenamefont {Leibfried}\ \emph
  {et~al.}(2003){\natexlab{a}}\citenamefont {Leibfried}, \citenamefont {Blatt},
  \citenamefont {Monroe},\ and\ \citenamefont {Wineland}}]{Leibfried}%
  \BibitemOpen
  \bibfield  {author} {\bibinfo {author} {\bibfnamefont {D.}~\bibnamefont
  {Leibfried}}, \bibinfo {author} {\bibfnamefont {R.}~\bibnamefont {Blatt}},
  \bibinfo {author} {\bibfnamefont {C.}~\bibnamefont {Monroe}}, \ and\ \bibinfo
  {author} {\bibfnamefont {D.}~\bibnamefont {Wineland}},\ }\href@noop {}
  {\bibfield  {journal} {\bibinfo  {journal} {Rev. Mod. Phys.},\ }\textbf
  {\bibinfo {volume} {75}},\ \bibinfo {pages} {281} (\bibinfo {year}
  {2003}{\natexlab{a}})}\BibitemShut {NoStop}%
\bibitem [{\citenamefont {{Wallraff}}\ \emph {et~al.}(2004)\citenamefont
  {{Wallraff}}, \citenamefont {{Schuster}}, \citenamefont {{Blais}},
  \citenamefont {{Frunzio}}, \citenamefont {{Huang}}, \citenamefont {{Majer}},
  \citenamefont {{Kumar}}, \citenamefont {{Girvin}},\ and\ \citenamefont
  {{Schoelkopf}}}]{Wallraff04}%
  \BibitemOpen
  \bibfield  {author} {\bibinfo {author} {\bibfnamefont {A.}~\bibnamefont
  {{Wallraff}}}, \bibinfo {author} {\bibfnamefont {D.~I.}\ \bibnamefont
  {{Schuster}}}, \bibinfo {author} {\bibfnamefont {A.}~\bibnamefont {{Blais}}},
  \bibinfo {author} {\bibfnamefont {L.}~\bibnamefont {{Frunzio}}}, \bibinfo
  {author} {\bibfnamefont {R.-S.}\ \bibnamefont {{Huang}}}, \bibinfo {author}
  {\bibfnamefont {J.}~\bibnamefont {{Majer}}}, \bibinfo {author} {\bibfnamefont
  {S.}~\bibnamefont {{Kumar}}}, \bibinfo {author} {\bibfnamefont {S.~M.}\
  \bibnamefont {{Girvin}}}, \ and\ \bibinfo {author} {\bibfnamefont {R.~J.}\
  \bibnamefont {{Schoelkopf}}},\ }\Doi {10.1038/nature02851} {\bibfield
  {journal} {\bibinfo  {journal} {\nat},\ }\textbf {\bibinfo {volume} {431}},\
  \bibinfo {pages} {162} (\bibinfo {year} {2004})}\BibitemShut {NoStop}%
\bibitem [{\citenamefont {Gross}\ and\ \citenamefont
  {Haroche}(1982)}]{Haroche}%
  \BibitemOpen
  \bibfield  {author} {\bibinfo {author} {\bibfnamefont {M.}~\bibnamefont
  {Gross}}\ and\ \bibinfo {author} {\bibfnamefont {S.}~\bibnamefont
  {Haroche}},\ }\href@noop {} {\bibfield  {journal} {\bibinfo  {journal}
  {Physics Reports}} (\bibinfo {year} {1982})}\BibitemShut {NoStop}%
\bibitem [{\citenamefont {Louko}\ and\ \citenamefont {Schleich}(1999)}]{LoSch}%
  \BibitemOpen
  \bibfield  {author} {\bibinfo {author} {\bibfnamefont {J.}~\bibnamefont
  {Louko}}\ and\ \bibinfo {author} {\bibfnamefont {K.}~\bibnamefont
  {Schleich}},\ }\href@noop {} {\bibfield  {journal} {\bibinfo  {journal}
  {Classical and Quantum Gravity},\ }\textbf {\bibinfo {volume} {16}},\
  \bibinfo {pages} {2005} (\bibinfo {year} {1999})}\BibitemShut {NoStop}%
\bibitem [{\citenamefont {Misner}\ \emph {et~al.}(1973)\citenamefont {Misner},
  \citenamefont {Thorne},\ and\ \citenamefont {Wheeler}}]{gravitation}%
  \BibitemOpen
  \bibfield  {author} {\bibinfo {author} {\bibfnamefont {C.~W.}\ \bibnamefont
  {Misner}}, \bibinfo {author} {\bibfnamefont {K.~S.}\ \bibnamefont {Thorne}},
  \ and\ \bibinfo {author} {\bibfnamefont {J.~A.}\ \bibnamefont {Wheeler}},\
  }\href@noop {} {\emph {\bibinfo {title} {Gravitation}}}\ (\bibinfo
  {publisher} {{W. H. Freeman}},\ \bibinfo {year} {1973})\BibitemShut {NoStop}%
\bibitem [{\citenamefont {Mart\'\i{}n-Mart\'\i{}nez}\ \emph
  {et~al.}(2010)\citenamefont {Mart\'\i{}n-Mart\'\i{}nez}, \citenamefont
  {Garay},\ and\ \citenamefont {Le\'on}}]{Edu6}%
  \BibitemOpen
  \bibfield  {author} {\bibinfo {author} {\bibfnamefont {E.}~\bibnamefont
  {Mart\'\i{}n-Mart\'\i{}nez}}, \bibinfo {author} {\bibfnamefont {L.~J.}\
  \bibnamefont {Garay}}, \ and\ \bibinfo {author} {\bibfnamefont
  {J.}~\bibnamefont {Le\'on}},\ }\href@noop {} {\bibfield  {journal} {\bibinfo
  {journal} {Phys. Rev. D},\ }\textbf {\bibinfo {volume} {82}},\ \bibinfo
  {pages} {064006} (\bibinfo {year} {2010})}\BibitemShut {NoStop}%
\bibitem [{\citenamefont {{Schneider}}\ \emph {et~al.}(2011)\citenamefont
  {{Schneider}}, \citenamefont {{Porras}},\ and\ \citenamefont
  {{Schaetz}}}]{Schneider11}%
  \BibitemOpen
  \bibfield  {author} {\bibinfo {author} {\bibfnamefont {C.}~\bibnamefont
  {{Schneider}}}, \bibinfo {author} {\bibfnamefont {D.}~\bibnamefont
  {{Porras}}}, \ and\ \bibinfo {author} {\bibfnamefont {T.}~\bibnamefont
  {{Schaetz}}},\ }\href@noop {} {\bibfield  {journal} {\bibinfo  {journal}
  {ArXiv e-prints}} (\bibinfo {year} {2011})},\ \Eprint
  {http://arxiv.org/abs/1106.2597} {arXiv:1106.2597 [quant-ph]} \BibitemShut
  {NoStop}%
\bibitem [{\citenamefont {{Gerritsma}}\ \emph {et~al.}(2010)\citenamefont
  {{Gerritsma}}, \citenamefont {{Kirchmair}}, \citenamefont {{Z{\"a}hringer}},
  \citenamefont {{Solano}}, \citenamefont {{Blatt}},\ and\ \citenamefont
  {{Roos}}}]{naturekike}%
  \BibitemOpen
  \bibfield  {author} {\bibinfo {author} {\bibfnamefont {R.}~\bibnamefont
  {{Gerritsma}}}, \bibinfo {author} {\bibfnamefont {G.}~\bibnamefont
  {{Kirchmair}}}, \bibinfo {author} {\bibfnamefont {F.}~\bibnamefont
  {{Z{\"a}hringer}}}, \bibinfo {author} {\bibfnamefont {E.}~\bibnamefont
  {{Solano}}}, \bibinfo {author} {\bibfnamefont {R.}~\bibnamefont {{Blatt}}}, \
  and\ \bibinfo {author} {\bibfnamefont {C.~F.}\ \bibnamefont {{Roos}}},\ }\Doi
  {10.1038/nature08688} {\bibfield  {journal} {\bibinfo  {journal} {Nature},\
  }\textbf {\bibinfo {volume} {463}},\ \bibinfo {pages} {68} (\bibinfo {year}
  {2010})} \BibitemShut {NoStop}%
  \bibitem [{\citenamefont {Menicucci}\ \emph {et~al.}(2010)\citenamefont
  {Menicucci}, \citenamefont {Olson},\ and\ \citenamefont {Milburn}}]{Milburn2}%
  \BibitemOpen
  \bibfield  {author} {\bibinfo {author} {\bibfnamefont {P.~M.}\ \bibnamefont
  {Alsing}}, \bibinfo {author} {\bibfnamefont {J.~P.}\ \bibnamefont
  {Dowling}}, \ and\ \bibinfo {author} {\bibfnamefont {G.~J.}\ \bibnamefont
  {Milburn}},\ }\href@noop {} {\bibfield  {journal} {\bibinfo  {journal} {Phys. Rev. Let.},\ }\textbf {\bibinfo {volume} {94}},\ \bibinfo {pages}
  {220401} (\bibinfo {year} {2005})}\BibitemShut {NoStop}%
\bibitem [{\citenamefont {Menicucci}\ \emph {et~al.}(2010)\citenamefont
  {Menicucci}, \citenamefont {Olson},\ and\ \citenamefont {Milburn}}]{Milburn}%
  \BibitemOpen
  \bibfield  {author} {\bibinfo {author} {\bibfnamefont {N.~C.}\ \bibnamefont
  {Menicucci}}, \bibinfo {author} {\bibfnamefont {S.~J.}\ \bibnamefont
  {Olson}}, \ and\ \bibinfo {author} {\bibfnamefont {G.~J.}\ \bibnamefont
  {Milburn}},\ }\href@noop {} {\bibfield  {journal} {\bibinfo  {journal} {New
  Journal of Physics},\ }\textbf {\bibinfo {volume} {12}},\ \bibinfo {pages}
  {095019} (\bibinfo {year} {2010})}\BibitemShut {NoStop}%
\bibitem [{\citenamefont {Leibfried}\ \emph
  {et~al.}(2003){\natexlab{b}}\citenamefont {Leibfried}, \citenamefont {Blatt},
  \citenamefont {Monroe},\ and\ \citenamefont {Wineland}}]{Leibfried03}%
  \BibitemOpen
  \bibfield  {author} {\bibinfo {author} {\bibfnamefont {D.}~\bibnamefont
  {Leibfried}}, \bibinfo {author} {\bibfnamefont {R.}~\bibnamefont {Blatt}},
  \bibinfo {author} {\bibfnamefont {C.}~\bibnamefont {Monroe}}, \ and\ \bibinfo
  {author} {\bibfnamefont {D.}~\bibnamefont {Wineland}},\ }\Doi
  {10.1103/RevModPhys.75.281} {\bibfield  {journal} {\bibinfo  {journal} {Rev.
  Mod. Phys.},\ }\textbf {\bibinfo {volume} {75}},\ \bibinfo {pages} {281}
  (\bibinfo {year} {2003}{\natexlab{b}})}\BibitemShut {NoStop}%
\bibitem [{\citenamefont {Porras}\ \emph {et~al.}(2008)\citenamefont {Porras},
  \citenamefont {Marquardt}, \citenamefont {von Delft},\ and\ \citenamefont
  {Cirac}}]{Porras08}%
  \BibitemOpen
  \bibfield  {author} {\bibinfo {author} {\bibfnamefont {D.}~\bibnamefont
  {Porras}}, \bibinfo {author} {\bibfnamefont {F.}~\bibnamefont {Marquardt}},
  \bibinfo {author} {\bibfnamefont {J.}~\bibnamefont {von Delft}}, \ and\
  \bibinfo {author} {\bibfnamefont {J.~I.}\ \bibnamefont {Cirac}},\ }\Doi
  {10.1103/PhysRevA.78.010101} {\bibfield  {journal} {\bibinfo  {journal}
  {Phys. Rev. A},\ }\textbf {\bibinfo {volume} {78}},\ \bibinfo {pages}
  {010101} (\bibinfo {year} {2008})}\BibitemShut {NoStop}%
\bibitem [{\citenamefont {{Porras}}\ and\ \citenamefont {{Jos{\'e}
  Garc{\'{\i}}a-Ripoll}}(2011)}]{Porras11}%
  \BibitemOpen
  \bibfield  {author} {\bibinfo {author} {\bibfnamefont {D.}~\bibnamefont
  {{Porras}}}\ and\ \bibinfo {author} {\bibfnamefont {J.}~\bibnamefont
  {{Jos{\'e} Garc{\'{\i}}a-Ripoll}}},\ }\href@noop {} {\bibfield  {journal}
  {\bibinfo  {journal} {ArXiv e-prints}} (\bibinfo {year} {2011})},\ \Eprint
  {http://arxiv.org/abs/1107.2607} {arXiv:1107.2607 [quant-ph]} \BibitemShut
  {NoStop}%
\bibitem [{\citenamefont {{Zener}}(1932)}]{zener}%
  \BibitemOpen
  \bibfield  {author} {\bibinfo {author} {\bibfnamefont {C.}~\bibnamefont
  {{Zener}}},\ }\href@noop {} {\bibfield  {journal} {\bibinfo  {journal} {Royal
  Society of London Proceedings Series A},\ }\textbf {\bibinfo {volume}
  {137}},\ \bibinfo {pages} {696} (\bibinfo {year} {1932})}\BibitemShut
  {NoStop}%
\end{thebibliography}

\end{document}